%% file: main.tex
\documentclass[sigconf]{acmart}

\usepackage{amsmath,amsfonts} 
\usepackage{algorithmic}
\usepackage{graphicx}
\usepackage{textcomp}
\usepackage{fancyhdr}
\usepackage{xcolor}
\usepackage{adjustbox}
\usepackage{lipsum}
\usepackage{units}
\usepackage{hyperref}
\usepackage{comment} 

\AtBeginDocument{%
  \providecommand\BibTeX{{%
    \normalfont B\kern-0.5em{\scshape i\kern-0.25em b}\kern-0.8em\TeX}}}

\usepackage{color, colortbl}
\definecolor{Gray}{gray}{0.9}

\setcopyright{rightsretained}
\copyrightyear{2021}
\acmYear{2021}
\acmConference[ICPE '21 Companion]{Companion of the 2021 ACM/SPEC International Conference on Performance Engineering}{April 19--23, 2021}{Virtual Event, France}
\acmBooktitle{Companion of the 2021 ACM/SPEC International Conference on Performance Engineering (ICPE '21 Companion), April 19--23, 2021, Virtual Event, France}\acmDOI{10.1145/3447545.3451183}
\acmISBN{978-1-4503-8331-8/21/04}
\begin{document}

\title{10 Years Later: Cloud Computing is Closing the Performance Gap}

\author{Giulia Guidi}
\affiliation{%
    \institution{\fontsize{9}{11}\selectfont University of California, Berkeley}
    \institution{Lawrence Berkeley National Laboratory}
    \city{Berkeley}
    \state{California}
    \country{USA}
}

\author{Marquita Ellis}
\affiliation{%
    \institution{\fontsize{9}{11}\selectfont University of California, Berkeley}
    \institution{Lawrence Berkeley National Laboratory}
    \city{Berkeley}
    \state{California}
    \country{USA}
}

\author{Ayd{\i}n Bulu\c{c}}
\affiliation{%
    \institution{\fontsize{9}{11}\selectfont University of California, Berkeley}
    \institution{Lawrence Berkeley National Laboratory}
    \city{Berkeley}
    \state{California}
    \country{USA}
}

\author{Katherine Yelick}
\affiliation{%
    \institution{\fontsize{9}{11}\selectfont University of California, Berkeley}
    \institution{Lawrence Berkeley National Laboratory}
    \city{Berkeley}
    \state{California}
    \country{USA}
}

\author{David Culler}
\affiliation{%
    \institution{\fontsize{9}{11}\selectfont University of California, Berkeley}
    \institution{Google, Inc.}
    \city{Berkeley}
    \state{California}
    \country{USA}
}

\renewcommand{\shortauthors}{Guidi et al.}

\begin{abstract}
    Can cloud computing infrastructures provide HPC-competitive performance for scientific applications broadly?
    Despite prolific related literature, this question remains open. Answers are crucial for designing future systems and democratizing high-performance computing. 
    We present a multi-level approach to investigate the performance gap between HPC and cloud computing, isolating different variables that contribute to this gap. Our experiments are divided into (i) hardware and system microbenchmarks and (ii) user application proxies. 
    The results show that 
    today's high-end cloud computing can deliver HPC-competitive performance not only for computationally intensive applications, but also for memory- and communication-intensive applications
    -- at least at modest scales -- thanks to the high-speed memory systems and interconnects and dedicated batch scheduling now available on some cloud platforms.
\end{abstract}


\settopmatter{printfolios=true}
\maketitle

\input{sec/introduction}
\input{sec/methodology}
\input{sec/microbenchmark}
\input{sec/application}
\input{sec/conclusions}
\begin{acks}
This work is supported by the Advanced Scientific Computing Research (ASCR) program within the Office of Science of the DOE under contract number DE-AC02-05CH11231. We used resources of the NERSC supported by the Office of Science of the DOE under Contract No. DEAC02-05CH11231. This research was also supported by the Exascale Computing Project (17-SC-20-SC), a collaborative effort of the U.S. Department of Energy Office of Science and the National Nuclear Security Administration.
AWS Cloud Credits provided through the AWS Cloud Credits for Research program.
Thanks to Rohan Bavishi, Rohan Padhye, Neesha Zerin, and James Demmel for useful suggestions and valuable discussions.
\end{acks}

\bibliographystyle{ACM-Reference-Format}
\bibliography{sample-base}


\end{document}

%% file: sec/introduction.tex
\section{Introduction}\label{sec:introdution}

The benefit of high-performance computing for scientific research has grown rapidly, beyond traditional simulation problems to data analysis in light sources, cosmology, genomics, particle physics, and more~\cite{yelick2020parallelism,alexander2020exascale}.
Given the vast amounts of data and/or computation involved in such applications, they can require the full computing power and memory of high performance computing (HPC) systems.
Cloud computing~\cite{fox2009above,buyya2009cloud,mell2011nist} is gaining popularity among scientists as an alternative to HPC for a wide range of sciences such as physics, bioinformatics, cosmology, and climate research~\cite{evangelinos2008cloud,mushtaq2017sparkga}.

There are many efforts in the literature to measure the performance of scientific applications in the cloud.
The lack of a low-latency network has been consistently identified as the main bottleneck~\cite{gupta2014evaluating,netto2018hpc,yelick2011magellan,ellis2017performance,ellis2019dibella}.
These studies have shown that the cloud delivers competitive performance for HPC applications with minimal communication and I/O, but significantly underperforms for memory- and communication-intensive workloads.
Virtualization overhead has also been identified as a performance-limiting factor, but studies do not generally agree on its impact. 
He et al.~\cite{he2010case} (2010) concluded that virtualization technology has no significant performance overhead, while the results in the Magellan report~\cite{yelick2011magellan} (2011) and by Gupta et al.~\cite{gupta2014evaluating} (2014) show that virtualization overhead along with slow network is one of the major limitations of the cloud. 
Performance variability due to resource sharing and the lack of tools for using and managing cloud environments--such as batch scheduling and base images--have also further limited the competitiveness of the cloud for scientific computing~\cite{yelick2011magellan, gupta2014evaluating, netto2018hpc}.

Understanding the nature of the gap between HPC and cloud systems and whether the results from the literature are still valid today is critical for guiding future system design and running scientific applications efficiently in the cloud.
Furthermore, the number of users that supercomputing facilities can support is limited, and the deployment of any new supercomputer is a multi-year multi-million dollar investment.
Closing the performance gap between HPC and the cloud would give many more scientists access to adequate computing resources.

Our work measures the performance of HPC-oriented codes on both cloud and HPC platforms. 
Building on previous literature, we investigate whether the findings apply to today's cloud platforms and isolate the contribution of different variables to the HPC cloud performance gap.
Our results show that cloud platforms with similar processors and networks can achieve HPC-competitive performance, not only for compute-intensive applications, but also for communication-intensive applications.
At small and medium scales, modern cloud computing has overcome one of its main limitations by providing higher-speed memory and interconnects for HPC-oriented instances.
Cost models, job wait times, availability of pre-installed software, and other usability factors are also relevant to evaluating HPC-cloud competitiveness, but are out of scope here.

%% file: sec/methodology.tex
\section{Background}\label{sec:bkg}

High Performance Computing (HPC) and cloud computing differ in their original purpose as well as their economic objectives and access policies.
HPC systems were designed to deliver high performance for dedicated scientific computing, while cloud computing made networked hardware and software available for general use.

The differences in their economic objectives and access policies inevitably affect scheduling, hardware selection, and software configuration decisions. 
HPC systems are typically operated by a non-profit organization (university or national laboratory), funded by a government agency, and allocated to a particular research community. 
These systems have very high utilization (over 90\%) with non-trivial wait times for users;  they support very large-scale computations with homogeneous hardware that undergoes major upgrades every few years.
In contrast, cloud systems are built for profit, configured to meet market demand, and operated at lower utilization rates to ensure little or no wait time. 
Cloud resources are upgraded continuously and incrementally, leading to rapid access to new technologies, but also to heterogeneity within the cloud.

Researchers running scientific applications in the cloud can access instances with low-latency networks to achieve performance competitive to HPC~\cite{he2010case}. 
However, cloud heterogeneity can limit what is available within an HPC cloud offering, e.g., it may be more difficult for a user to obtain a large number of high-performance instances.
In the cloud, users can easily customize their environment without administrative overhead and quickly provision additional resources to solve large problems~\cite{yelick2011magellan}.
HPC platforms offer limited support for on-demand self-service~\cite{yelick2011magellan}, but they do offer important features such as resource pooling and broad network access.

 The basic business model differences between cloud and HPC persist today and lead to complex cost trade-offs that are beyond the scope of this paper.  However, the growing commercial interest in problems such as large-scale machine learning training have led to changes in cloud configurations.  This has increased the popularity of HPC-as-a-Service in the cloud and has in turn resurfaced questions about use of the cloud for modest scale parallel scientific applications.  

\subsection{Low-Level Benchmark}

A common approach to comparing the performance of computer systems is to use low-level benchmarks~\cite{tsouloupas2006characterization,alam2006characterization}.
Here, we focus on the investigation of processor, memory, and network performance.

{\bf Processor.} 
Considering a multi-core processor, we refer to it as a \emph{node}, where a \emph{core} is the basic execution unit in the system.
The number of nodes is later denoted $P$.
Here we use the shared memory version of the LINPACK benchmark~\cite{dongarra2003linpack} to compare the floating point performance of the systems under consideration.
LINPACK measures performance as the number of $64$-bit floating point operations a computer can perform per second (FLOPS).
The performance when running an actual application is likely to be lower than the performance achieved by the LINPACK benchmark.

{\bf Memory Hierarchy.}
CacheBench~\cite{LLCbench23:online} measures the performance of the local memory hierarchy.
It computes a number of operations – \emph{read, write, read/modify/write}, \emph{memset}, and \emph{memcpy} – varying the underlying array size, thereby revealing the performance of the cache.
Operations run for 2 seconds and the average bandwidth (MB/s) is reported. 
Here we focus on \emph{memcpy}.

{\bf Memory Bandwidth.}
To measure the maximum memory bandwidth of our systems, we use the STREAM benchmark~\cite{mccalpin1995stream}, which performs four vector operations: \emph{copy}, \emph{scale}, \emph{sum}, and \emph{triad}.
STREAM requires that (a) each array is at least four times the size of the cache memory, and (b) the size is such that the ``timing calibration'' output by the program is at least 20 clock ticks.
STREAM provides the best possible memory system bandwidth.

{\bf Inter-Node Communication.}
Following standard practice~\cite{he2010case}, we use a subset of MPI operations to measure the inter-node communication performance of our systems.
Specifically, we use \texttt{MPI\_Send-} \texttt{recv} and \texttt{MPI\_Alltoall} to measure point-to-point and collective latency and bandwidth using the OSU microbenchmarks~\cite{panda2018osu}.

\subsection{User Application}

Besides comparing HPC and cloud systems on a subset of MPI collectives, we select two representative user applications from scientific computing as benchmarks: an N-Body simulation written in \texttt{C++} and a Fast Fourier Transform (FFT), written in \texttt{C}.
N-Body is a computationally intensive application, while the FFT is more communication intensive~\cite{woo1995splash,asanovic2006landscape,colella2004defining}.

An N-Body simulation models a dynamic system of particles, usually under the influence of physical forces, such as gravity~\cite{ida1992n}.
It is a common computation in physics, astronomy, and biology.
The naive solution computes the forces acting on the particles by iterating through each pair of particles, resulting in a complexity of $O(n^2)$, where $n$ is the number of particles.
In our implementation we consider the density of the particles to be sufficiently low so that a linear time solution can be achieved with $n$ particles.
 
The FFT calculates the discrete Fourier transform (DFT) of a sequence or its inverse (IDFT). 
In Fourier analysis, a signal is transformed from its original domain (often time or space) to a frequency domain representation and vice versa.
As a benchmark for the FFT, we use the implementation of Frigo and Johnson~\cite{frigo1997fastest,frigo1998fftw,frigo1999fast}, Fast Fourier Transform in the West (FFTW).

%% file: sec/microbenchmark.tex
\section{Empirical Methodology}\label{sec:exp}

Processor, memory, network, application and programming model, and system age are all variables that affect performance.
Here, we measure the performance gap by isolating the contribution of the different variables by dividing our experiments into two categories: (i) hardware and system and (ii) user application.

First, we isolate the contribution of processor and memory to identify similarities or significant differences in in-node performance.
Then, we investigate the contribution of the inter-processor network by measuring the latency and bandwidth of communication primitives between machines.
Finally, we study performance of HPC and cloud computing from an application perspective.
Unless differently noted, the results reported in this paper represent the average value across 10 runs.
 
To this end, we use two metrics to characterize our applications: hardware events and the communication to computation ratio ($Cm/Cp$).
The $Cm/Cp$ ratio is defined as communication time divided by computation time for a given execution of a parallel application on a given parallel machine with explicit communication~\cite{crovella1992using}.
Both metrics can help interpret the potential performance gap between HPC and cloud systems.

\subsection{Experimental Setting}\label{sec:methodology}

\input{sec/tab/tab-spec}

Our experiments are conducted on the Intel Xeon ``Haswell'' (Cori Haswell) and Intel Xeon Phi ``Knight's Landing'' (KNL) partitions (Cori KNL) of the Cori Cray XC40 HPC system at NERSC, an Amazon Web Services (AWS) commodity cluster with r5dn.16xlarge (R5) instances (optimized for memory-intensive workloads), and one with AWS c5.18xlarge (C5) instances (optimized for compute- intensive workloads).
Details for each instance are listed in Table~\ref{tab:spec}.

We chose these four platforms because of their easy availability and the diversity of architectures.
In particular, we selected the two AWS instances to represent two extremes of the AWS catalog (memory-optimized versus compute-optimized) and selected these two instances because they allowed us to allocate multiple nodes in the same placement group in a reasonable amount of time.
AWS clusters run as \emph{dedicated instances} to reduce the potential performance slowdown from sharing resources, and use Slurm as the workload manager~\cite{yoo2003slurm}.
The need for tools to simplify the use of cloud environments and better software stacks for clouds has been noted in past literature~\cite{yelick2011magellan}.
We use AWS ParallelCluster to provision and manage AWS clusters.
It automatically sets up the required compute resources and shared file system in about five to ten minutes in our experience.
AWS also provides a collection of Amazon Machine Images (AMIs) installed with libraries and software such as MPI, BLAS, and TensorFlow.
Notably, AWS ParallelCluster provides support for several schedulers, such as SGE and Torque (which will both be discontinued at the end of 2021), as well as Slurm and the in-house AWS Batch, which currently has limited support for GPU jobs. 
We used Slurm for consistency across AWS and Cori systems.

Cori Haswell and KNL also use Slurm as workload manager.
Cori has the Cray Aries ``Dragonfly'' topology for its interconnect~\cite{Intercon7:online}.
AWS does not disclose details about the underlying interconnect topology, except for an expected injection bandwidth (Table~\ref{tab:spec}), and that the AWS C5 instances use Amazon in-house Elastic Fabric Adapter (EFA)~\cite{AmazonEC22:online, ElasticF32:online} as network interface. 
The AWS cluster instances belong to the same placement group; the login node and the compute nodes belong to two different subnets.
A subnet is a logically visible subdivision of an IP network.
The subnetwork of compute nodes is private and has no access to the Internet.

\subsection{A Hardware and System View}\label{sec:microbench}

\begin{figure}[t]
\centering
\includegraphics[width=\columnwidth]{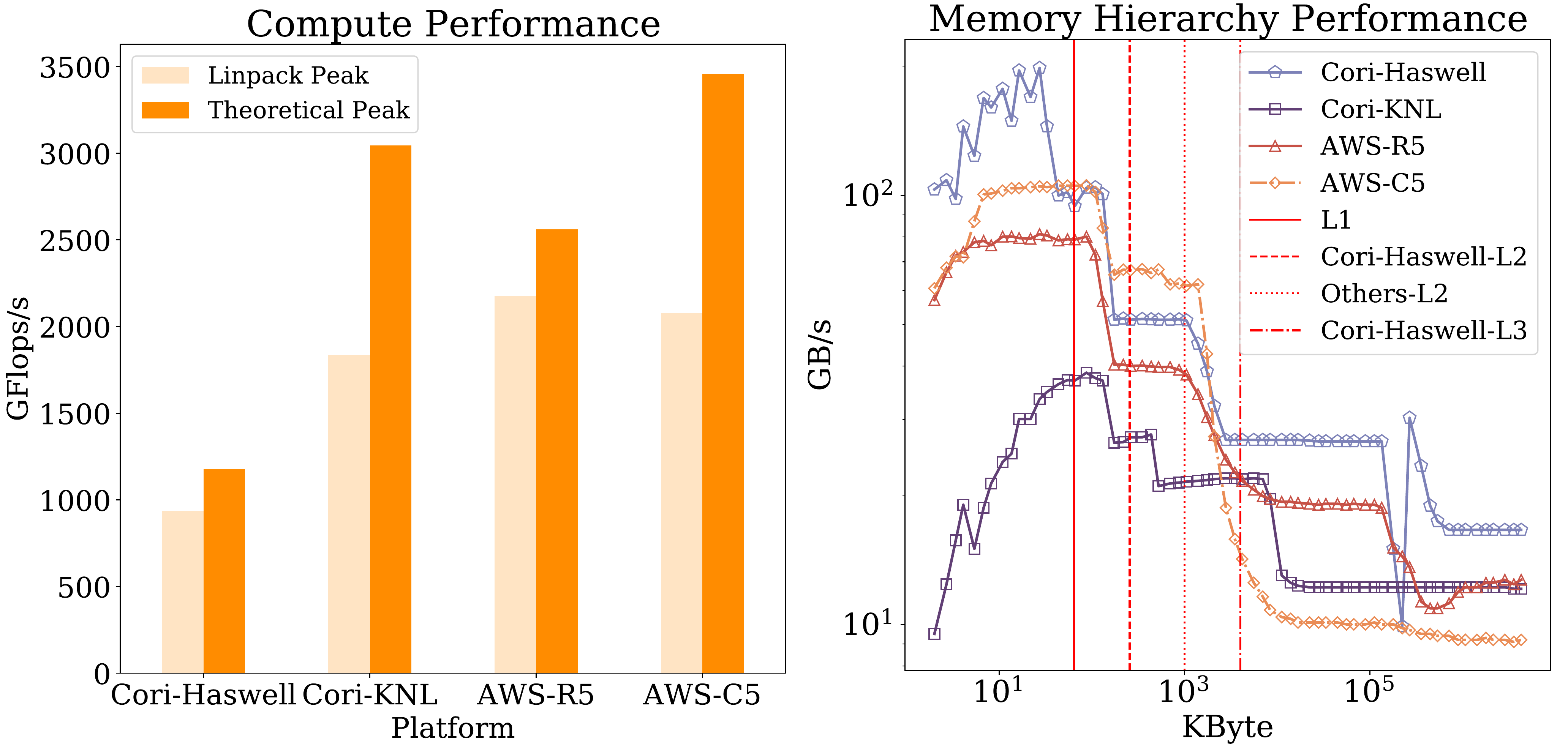}
\caption{LINPACK peak performance compared to the theoretical peak (left) using one node and all available cores per node (Table~\ref{tab:spec}). 
CacheBench \texttt{memcpy()} benchmark (right) using a single core and reporting the median of 10 runs.
}
\vspace{-1em}
\label{fig:linpackmem}
\end{figure}

Given different in-node configurations, we first investigate performance using a microbenchmarking approach.

{\bf Processor.}\label{sec:sys1}
Figure~\ref{fig:linpackmem} compares the LINPACK peak performance (left) with the theoretical peak performance (right) for each platform and also allows cross-platform comparison.
Cori Haswell and AWS R5 achieve peak performance significantly closer to their theoretical peak than the other two machines.
Closing the gap between theoretical peak and LINPACK peak on Cori KNL is notoriously difficult; achieving such progress requires a significant optimization effort for applications in general~\cite{barnes2016evaluating,RooflineKNL2016,MultinodeKNL2018}.
Cori KNL achieves about 350 GFlops/second more in our benchmark than the number reported in the Top500~\cite{CoriCray93:online}.
This discrepancy could be due to different implementations of the LINPACK benchmark, since we use the Intel Math Kernel Library benchmark package.
Further profiling of AWS C5 with VTune~\cite{reinders2005vtune} revealed a relatively low core utilization for this platform, which could explain the large gap between theoretical and achieved peak.

The cloud instances perform best in absolute terms.
AWS R5 and C5 instances are equipped with newer hardware than Cori systems; this may explain the greater processing power.
It is noteworthy that the elastic nature of cloud computing $-$ as opposed to multi-year projects to develop and install supercomputers $-$ offers the potential for rapid hardware turnaround.

{\bf Memory Hierarchy.}\label{sec:sys2}
Figure~\ref{fig:linpackmem} shows the results for the Cache-Bench benchmark (on the right) and illustrates the performance of the cache hierarchy for our four machines.
For each platform and size, we ran the benchmark 10 times and report the median; there is little variance among different runs for a given size and platform.

\input{sec/tab/tab-stream}

Cori Haswell has the best performance for L1 (which is the same size on all machines).
The L2 performance of Cori Haswell and AWS C5 are comparable, while the performance of Cori Haswell falls below that of the AWS C5 platform below its second cache level.
AWS C5 achieves better performance than Cori Haswell as long as the data fits into its L2 cache, and its performance falls below Cori Haswell when it enters the third cache level as expected because Cori Haswell has a larger L3 cache.

Considering the data in Table~\ref{tab:spec}, one would expect a higher bandwidth for AWS R5 and Cori KNL given their larger L2.
However, the way the caches are shared between the cores and cache associativity could affect overall memory throughput.
For Cori Haswell, L2 is private to each core, while for Cori KNL it is shared by two cores.
Cori KNL has two cache levels instead of three like the other machines.
Cori Haswell's cache is $8$-way associative, while Cori KNL has a direct mapped cache.
This direct mapping reduces cache management complexity, but can significantly increase cache thrashing, resulting in a high rate of cache misses and main memory accesses~\cite{KNLCache1:online}.

Looking only at these single core results, one might suspect that the virtualization overhead could prevent cloud instances from fully exploiting the potential of their caches.
However, our results, which measure the performance of the whole memory system, discourage this hypothesis, as shown in the next microbenchmark.

{\bf Memory Bandwidth.}\label{sec:sys3}
To measure memory bandwidth when data does not fit in the system cache, we run the STREAM benchmark~\cite{mccalpin1995stream}. 
The results in Table~\ref{tab:stream} show that the performance difference between Cori Haswell and AWS R5 and C5 (Figure~\ref{fig:linpackmem}) is reversed in favor of the cloud clusters when all available cores are used if the data does not fit in the platforms' caches.

Cori KNL has the higher memory bandwidth thanks to its on-chip multi-channel DRAM (MCDRAM) chip of $16$GB.
Looking at platforms without on-chip memory, cloud instances show a significantly higher memory bandwidth than the corresponding HPC platform.
System age and newer cloud hardware can explain this performance.
These results suggest that a faster hardware turnaround time could benefit not only computationally intensive applications, but also data-intensive applications.
In addition, these results discredit the hypothesis that virtualization overhead is a major limitation of today's cloud computing.

{\bf Inter-Node Communication.}\label{sec:sys4}
To study network performance, we measure bandwidth and latency in a multinode setting.
In our experiments, we use \texttt{openmpi-4.0.2} as the MPI implementation.
For Cori Haswell and KNL, we ran the benchmark suite with both \texttt{openmpi-4.0.2} and the default \texttt{cray-mpi}.
They provided similar performance, and we decided to report only the results for OpenMPI for clarity and consistency with the cloud instances.

\begin{figure}[t]
\centering
\includegraphics[width=\columnwidth]{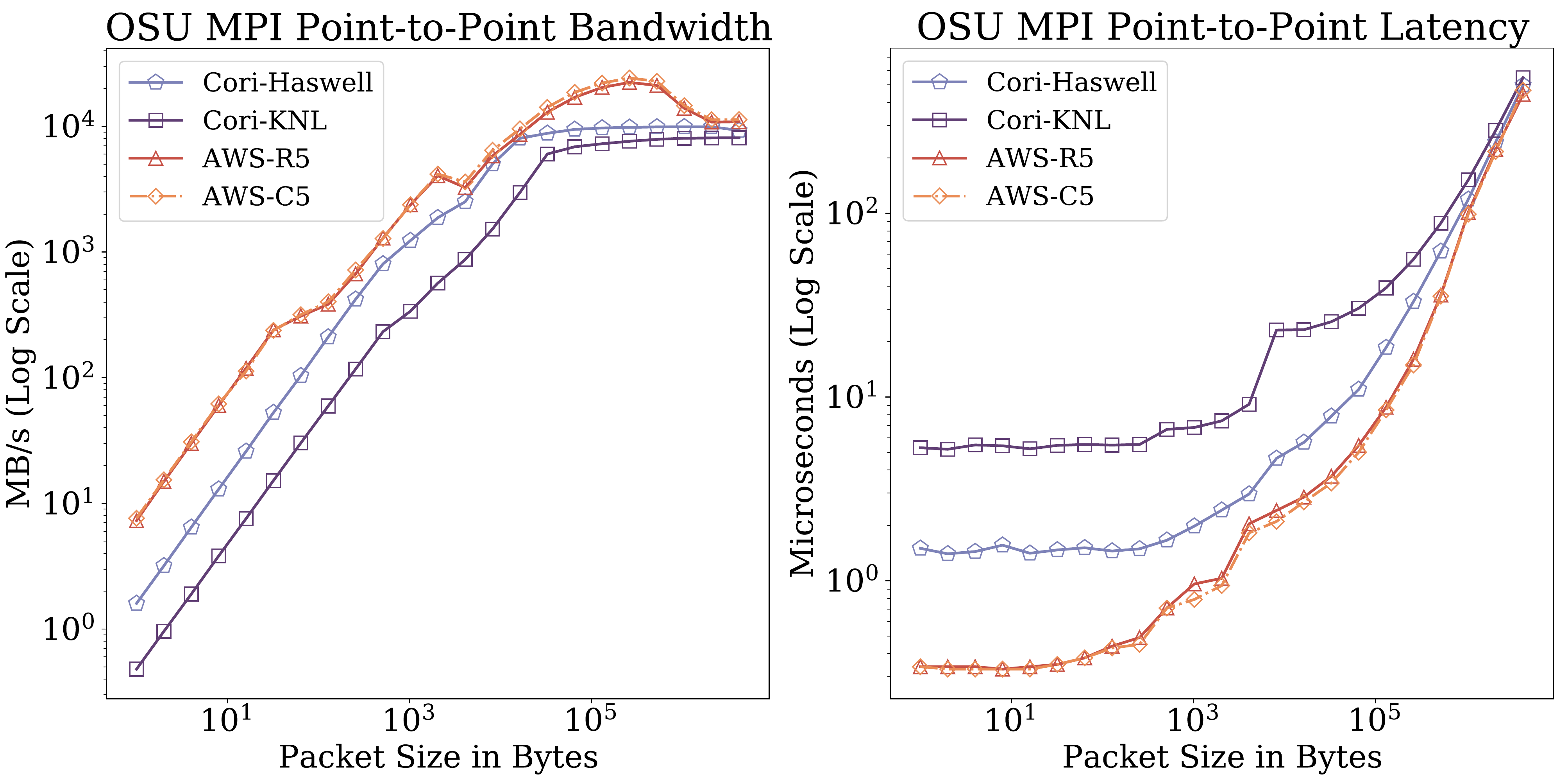}
\caption{OSU MPI microbenchmark injection bandwidth (left) and point-to-point latency (right) in log-log scale. Using two nodes with one process per node on Cori Haswell, Cori KNL, AWS R5, and AWS C5.
}
\vspace{-1em}
\label{fig:osuptp}
\end{figure}

Figure~\ref{fig:osuptp} uses one process per node to show point-to-point bandwidth (left) and latency (right).
Our results show peak bandwidth of about $86$ Gbit/s for AWS R5 and $90$ Gbit/s for AWS C5, while Cori Haswell and Cori KNL show peak bandwidths of $74$ and $64$ Gbit/s, respectively.
Considering that the two Cori systems share the same network, one would expect the same network performance, however, their performance in Figure~\ref{fig:osuptp} are significantly different.
This difference can be attributed to the overhead of MPI function calls, which are expensive and penalise lower frequency Cori KNL cores that cannot match the performance of Cori Haswell nodes.
Our results are consistent with those presented by GASNet~\cite{GASNetEX37:online}.
\begin{figure}[t]
\centering
\includegraphics[width=\columnwidth]{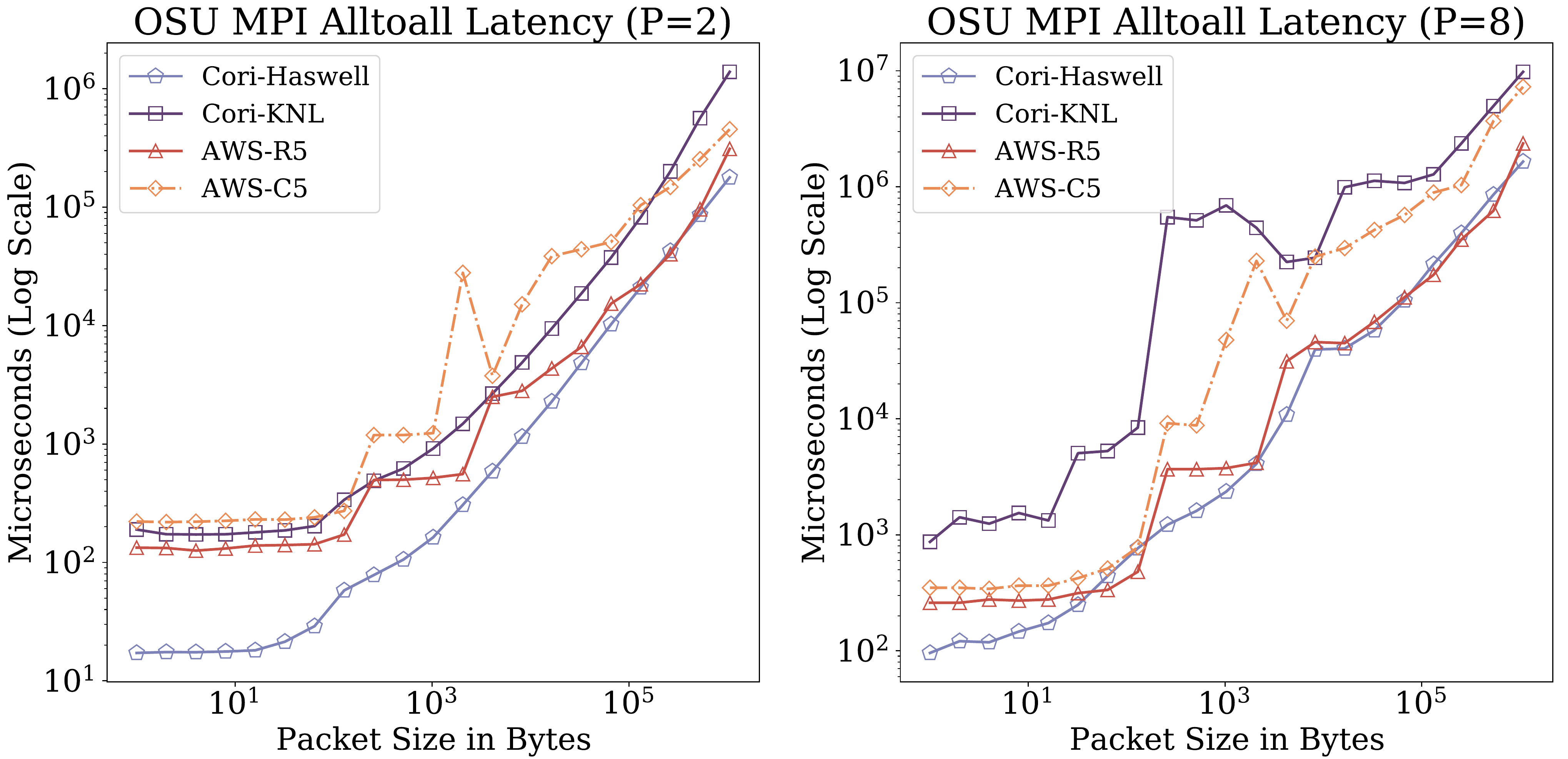}
\caption{OSU MPI microbenchmark \texttt{MPI\_Alltoall} latency on two nodes (left) and on eight nodes (right) in log-log scale. On Cori Haswell, AWS R5 and C5 we use 32 processes per node, while on Cori KNL we use 64 processes per node.}
\label{fig:osu-alltoall}
\end{figure}

Cloud instances outperform HPC systems in both bandwidth and latency.
Until recently, the lack of a low-latency network has been consistently identified as the main bottleneck of cloud computing for scientific applications~\cite{yelick2011magellan, gupta2014evaluating, netto2018hpc}.
Our results show that modern cloud computing has made significant advances in networking technology that provide cloud instances with HPC-competitive network performance. 

Figure~\ref{fig:osu-alltoall} shows the \texttt{MPI\_Alltoall} latency on $P{=}2,8$.
For small message sizes, Cori Haswell dominates the other platforms on two nodes; the gap decreases as the number of nodes is increased.
The differences between Cori Haswell and KNL are due to the cost of MPI calls on the two different processors.
Also, the different number of processes per node in this experiment illustrates the difference between the two Cori systems.
On two nodes, the gap decreases as the message size increases, especially when comparing Cori Haswell and AWS R5, whose performance almost overlaps at large message sizes.
AWS R5 shows similar performance to Cori Haswell on eight nodes, except for small message sizes.
Looking only at the historical results, one would expect the cloud instances to lose performance and the gap to grow as the number of nodes increases.
On the contrary, our results show significant improvements, so one can expect better performance scaling as the number of nodes increases.
AWS R5 performs as we would expect given its performance in the previous microbenchmark, while AWS C5 is far from Cori Haswell.
Its advertised network bandwidth is about $3\times$ lower than Cori Haswell and since it is a compute-optimized instance, we suspect it may suffer from network contention.

\input{sec/tab/tab-nbody}
\input{sec/tab/tab-fft}

Our results suggest that the place we would expect HPC to retain an advantage is in applications with many small messages.
Algorithmic techniques, however, typically try to avoid this situation.
These results have important implications for communication-intensive applications that have not historically benefited from cloud computing due to their bandwidth requirements.

%% file: sec/tab/tab-spec.tex
\begin{table*}[t]
\caption{Details of the evaluated machines: name, system age in years, number of cores per node, processor frequency, theoretical peak performance (GFlops/s) per node, processor, memory, advertised injection bandwidth (Gigabits/s), and caches sizes. $^\dagger$Custom model for Amazon AWS. KNL's L2 is shared between two cores.
$^*$Advertised user-process injection bandwidth~\cite{CrayXCSe62:online}.
}
\begin{adjustbox}{max width=\textwidth}{
\centering
\begin{tabular}{|l|c|c|c|c|l|r|r|r|r|r|r|}
\hline
\textbf{Platform} & \textbf{Age} & \textbf{Core/Node} & \textbf{Frequency (GHz)} & \textbf{Theoretical Peak (GFlops/Node)} &
\textbf{Processor} & \textbf{Memory (GiB)} & \textbf{Network (Gbps)} & \textbf{L1}\hspace{.65em} & \textbf{L2}\hspace{.65em}  & \textbf{L3}\hspace{.65em}  \\ 
\hline
\hline
Cori Haswell & 4 & 32  &  2.3 & 1,177 
& Xeon E5-2698V3 & 120          & $^*$82  & 64KB & 256KB &    40MB     \\
\rowcolor{Gray}
Cori KNL     & 4 & 68  & 1.4  & 3,046  
& Xeon Phi 7250        & 90           & $^*$82  & 64KB & 1MB &       -   \\
AWS r5dn.16xlarge  & 1 & 32  & 2.5  & 2,560 
& Xeon Platinum 8259CL          & 512      & 75    & 64KB & 1MB &   36MB      \\
\rowcolor{Gray}
AWS c5.18xlarge    & 1 & 36  & 3.0  & 3,456 
& Xeon Platinum 8124M$^\dagger$ & 144  & 25  & 64KB & 1MB &   25MB  \\
\hline
\end{tabular}
}
\end{adjustbox}
\label{tab:spec}
\end{table*}

%% file: sec/tab/tab-stream.tex
\begin{table}[t]
\centering
\caption{STREAM benchmark: as many OpenMP threads as the number of physical cores per node (top) and one thread (bottom), 8 bytes per array element, array size = 120000000 (elements), offset = 0 (elements), memory per array = 915.5 MiB, total memory required = 2746.6 MiB. The best time for each kernel over 10 runs (excluding the first iteration) is used to compute the bandwidth. Results in GB/s~\cite{mccalpin1995stream}.}
\label{tab:stream}
\begin{adjustbox}{max width=\columnwidth}{
\begin{tabular}{|l|c|r|r|r|r|}
\hline
\textbf{Platform} & \textbf{Threads} & \textbf{Copy} & \textbf{Scale} & \textbf{Add}\hspace{0.3em} & \textbf{Triad} \\ 
\hline
\hline
Cori Haswell  & 32 & 56.6 & 43.6  & 49.4  &49.7\\
\rowcolor{Gray}
Cori KNL  & 64 & 247.9  & 250.3  & 257.1 & 260.0 \\
AWS r5dn.16xlarge & 32 & 181.9 & 127.6 & 143.9 & 144.9 \\
\rowcolor{Gray}
AWS c5.18xlarge & 36 & 135.7 & 106.9 & 120.4 & 120.3 \\
\hline
\hline
Cori Haswell  & 1 & 18.0 & 11.3  & 12.6 & 12.6\\
\rowcolor{Gray}
Cori KNL  &  1 & 12.1 &  6.8   & 8.4 & 7.4 \\
AWS r5dn.16xlarge & 1 & 11.1 & 12.5 & 13.2 & 13.1 \\
\rowcolor{Gray}
AWS c5.18xlarge & 1 & 11.0 & 12.6 & 13.5 & 13.6\\
\hline
\end{tabular}
}
\end{adjustbox}
\vspace{-1em}
\end{table}

%% file: sec/tab/tab-nbody.tex
\begin{table}[t]
\centering
\caption{
Characterization of n-body using \texttt{perf} run on a single core. Page size = $4$KB, problem size: $1$M.}
\vspace{-1em}
\label{tab:app-nbodyc}
\begin{adjustbox}{max width=\columnwidth}{
\begin{tabular}{|l|r|r|r|r|r|}
\hline
\textbf{Platform} &  \textbf{Instruction (G)} & \textbf{Page Fault (K)} & \textbf{Cache Miss (M)} 
& \textbf{Time (s)} \\
\hline
\hline
Cori Haswell  & 414.7 & 367.2 &11,347.8 
&  461.7 \\
\rowcolor{Gray}
Cori KNL &415.4 &367.4 &11,220.1 
& 1,736.5\\
AWS r5dn.16xlarge   & - & 367.2 & - 
&486.9 \\
\rowcolor{Gray}
AWS c5.18xlarge  & 427.2 & 367.2 & 21,457.4 
& 480.6 \\
\hline
\end{tabular}
}
\end{adjustbox}
\end{table}

%% file: sec/tab/tab-fft.tex
\begin{table}[t]
\centering
\caption{
Characterization of FFT using \texttt{perf} run on a single core. Page size = $4$KB, problem size: $50$K.}
\vspace{-1em}
\label{tab:app-fft}
\begin{adjustbox}{max width=\columnwidth}{
\begin{tabular}{|l|r|r|r|r|r|}
\hline
\textbf{Platform} &  \textbf{Instruction (G)} & \textbf{Page Fault (K)} & \textbf{Cache Miss (M)} 
& \textbf{Time (s)} \\
\hline
\hline
Cori Haswell  & 782.1 &  9,766.8 & 871.5 
&  312.4 \\
\rowcolor{Gray}
Cori KNL  &784.9 & 9,766.8 & 20,915.0 
& 2,348.1 \\
AWS r5dn.16xlarge  & - & 9,767.5 & - 
& 303.3 \\
\rowcolor{Gray}
AWS c5.18xlarge   & 1,097.9 & 9,766.6 & 2,953.6 
&  335.8 \\
\hline
\end{tabular}
}
\end{adjustbox}
\vspace{-1em}
\end{table}

%% file: sec/application.tex
\subsection{A User-Application View}\label{sec:app}

In this section, we first measure and compare the serial runtime of the applications and analyze the single-core performance of the applications to better understand the runtime differences and similarities between the machines.
Then, we study the parallel performance of the applications in a multinode environment.

\subsubsection{Serial Performance}\label{sec:serialperf}

In Tables~\ref{tab:app-nbodyc} and~\ref{tab:app-fft}, we report the single core performance for the N-Body simulation and the FFT, respectively.
In both applications, Cori KNL has a significantly higher runtime than the other machines.
Its poor performance can be justified by the lower frequency of its processor and the poor performance of its memory system.
Cori KNL's clock speed is about half that of the other cores in the study, and it needs all 68 of them to compete with the (theoretical) GFlop rate of the other 32-36 core nodes.
Recall that the L2 caches on Cori KNL are shared by two cores, while they are private on the other machines.
In fact, the performance for FFT is relatively worse since it is a more memory intensive application than N-Body.
Cori Haswell and the two cloud instances show similar runtime for both applications.
Cloud instances have lower cache performance than Cori Haswell, while they have higher bandwidth when data can no longer fit in the cache. 
Since we study single-core performance here, the lower half of Table~\ref{tab:stream} shows that Cori Haswell and the AWS instances have comparable performance in the single-core STREAM benchmark.

Overall, these results are consistent with the results of our microbenchmarks and confirm that cloud virtualization overhead has decreased to a point where application performance is not significantly impacted.
As a result, cloud instances have comparable runtime to a HPC system for both applications.

\subsubsection{Workload Characterization}\label{sec:workloadchar}

Recall, when we measure the runtime of an application, we measure both the processor and the memory system.
Runtime alone is not enough to get a reasonable understanding of the variables that affect application performance.

Here, we extend our analysis by measuring the number of {\it page faults}, {\it instructions} and {\it cache misses} for each application on each platform and comparing the results.
A high rate of page swapping-in/out, cache misses, and a high number of instructions can significantly slow down applications~\cite{sherwood1999reducing,babka2009misses,lam1991cache}.
On all systems, these metrics are measured for a process on a single node using \texttt{perf}~\cite{weaver2013linux}.
Cache misses and instructions are not available for AWS R5. 
In particular, it is not easy to get access to accurate hardware counters.
On HPC systems they typically require administrative privileges, while on cloud systems it can be difficult to separate the effects of virtualization and gain access to accurate metrics.

\begin{figure}[t]
\centering
\includegraphics[width=\columnwidth]{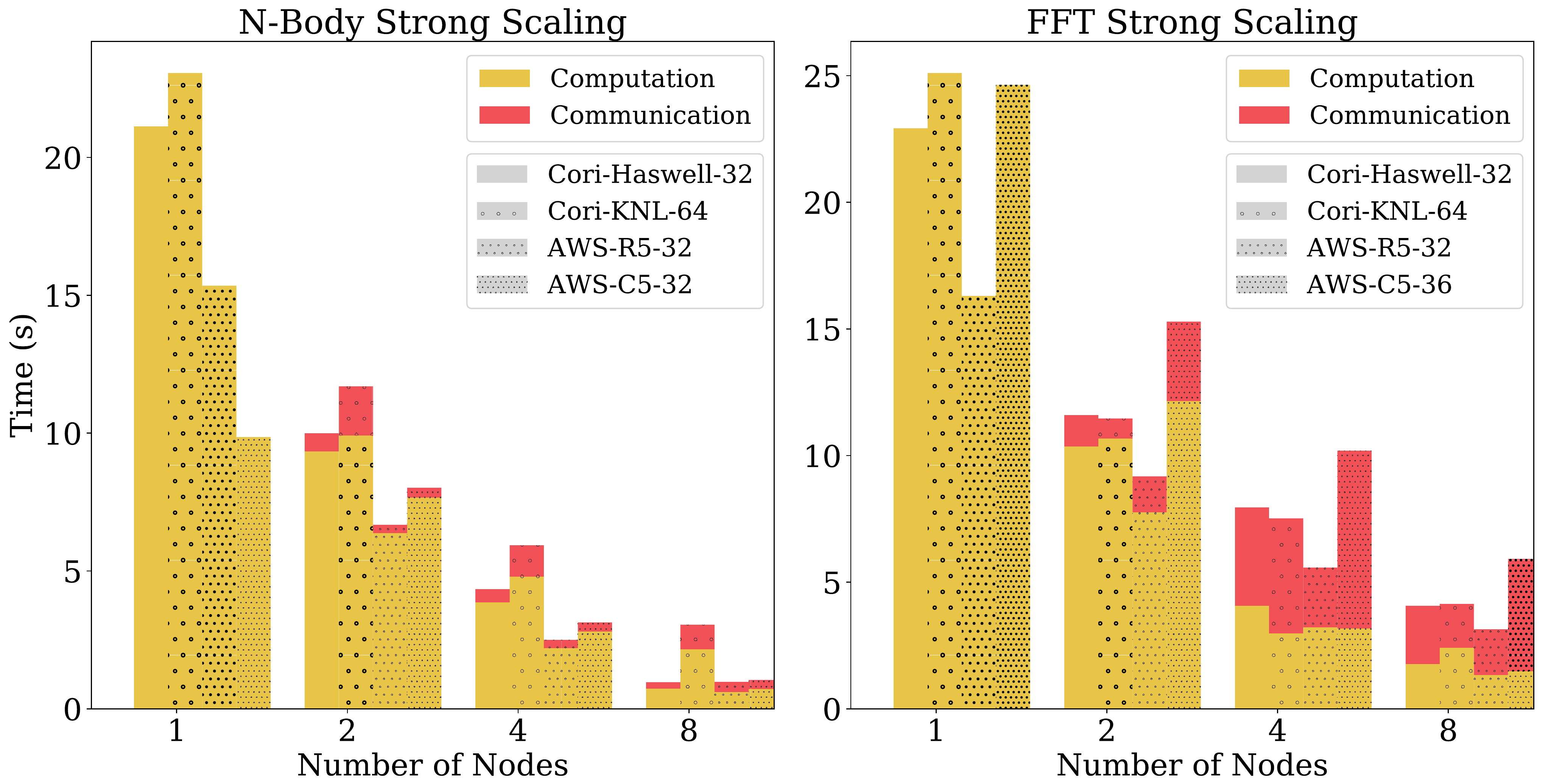}s
\caption{The N-Body strong scaling with $1$M particles (left) and the FFT strong scaling with $50$K points (right) across the machines.
The number next to the name in the legend indicates the number of processes per node.}
\vspace{-1em}
\label{fig:strongscaling}
\end{figure}

Tables~\ref{tab:app-nbodyc}$-$\ref{tab:app-fft} give the number of page faults on the machines for the N-Body simulation and the FFT.
The number of page faults is mostly the same and confirms the same behavior across the four machines.
Cori Haswell and Cori KNL automatically load a software package to increase the page size from $4$K to $2$M.
This setting was unloaded and disabled to allow a fair comparison between the four machines.
Similarly, the page size could have been increased on the AWS instances.
For simplicity, we chose to reduce the page size on Cori and do not expect this setting to change the overall trend of our results.
The only significant difference in the number of instructions is between the Cori systems and AWS C5 for the FFT.
This difference could explain the runtime difference between Cori Haswell and AWS C5, although it is not large.

Cache misses show a more relevant impact on performance than page faults and instructions.
The Cori systems have similar cache misses for N-Body simulation, while they show a significant gap for FFT.
Cori KNL's direct mapped cache significantly penalizes its performance for a memory-intensive application such as FFT.
AWS C5 has a larger number of cache misses than Cori Haswell for both applications.
This result, combined with AWS C5's slower L1 (Figure~\ref{fig:linpackmem}), suggests that cache misses are one of the variables contributing to the runtime difference between these two machines.

Our workload characterization reveals that cache misses and memory system performance have the largest impact on single-core performance.
Nevertheless, the resulting runtime differences are small, and our analysis shows comparable single-core performance between Cori Haswell and the cloud instances.

\subsubsection{Parallel Performance}\label{sec:parperf}

In examining parallel performance to highlight the effect of the network, we report the median of $10$ runs of the application for $P{=}1,2,4,8$.
Due to a limit on the number of instances we can create simultaneously, we were unable to get more than eight instances in the same placement group, which is critical for achieving low-latency network performance.
AWS support can increase this limit upon request.
Given the varying number of cores per node of our machines, we normalize our results and specify the configuration that provides the best performance for each platform.

{\bf N-Body Simulation.}\label{sec:synb}
Figure~\ref{fig:strongscaling} on the left illustrates the strong scaling performance across the machine and shows the runtime split in computation and communication. 
Our N-Body implementation uses a recursive doubling algorithm for particle exchange and therefore runs much faster with the power of two processes. 
As a result, all machines achieve their best performance with either $32$ or $64$ processes per node.

The N-Body simulation is computationally intensive and has a low $Cm/Cp$ ratio, suggesting a modest impact of the network on overall runtime.
Given the $Cm/Cp$ ratio and the serial performance of this application, we expect comparable runtimes between Cori Haswell and the two cloud instances.
Cori KNL also has comparable runtimes, while its $P{=}8$ scaling is significantly worse than the other three machines.
In particular, it uses twice as many processes per node as the other machines and runs at about half the frequency, with fewer cache levels and L2 caches shared by two cores.

The \texttt{MPI\_Alltoall} microbenchmark shows a significant difference between the two Cori systems.
Remember that the Cori systems use the same network; however, Cori KNL uses $64$ processes per node instead of $32$ and the MPI function calls overload the weaker KNL cores.
The gap is also significant between Cori Haswell and AWS C5 at any scale, while the gap between Cori Haswell and AWS R5 is mostly overlapping.

The \texttt{MPI\_Alltoall} gap between the two Cori systems is reflected in their performance in the N-Body simulation.
Therefore, one would expect AWS C5 to have a larger communication time. Nonetheless, the performance of AWS C5 is consistent with the assumption that when the $Cm/Cp$ ratio is low, the network has a limited impact on the overall application runtime.

Overall, AWS R5 is the fastest platform, with Cori Haswell and AWS C5 having the same performance at $P{=}8$.
The remarkable comeback of Cori Haswell might be due to fitting data into the larger L3 cache of Cori Haswell.
The N-Body simulation scales superlinearly $-$ on all machines except Cori KNL.
We are familiar with this implementation and know its superscaling behavior, which can be briefly explained as the ``cache effect'', meaning that as the number of nodes increases, more data fits into the cache.

Our results confirm that cloud computing can be more suitable than HPC systems for computationally intensive applications~\cite{yelick2011magellan} and that modern cloud computing can provide competitive network performance to HPC.

{\bf Fast Fourier Transform.}\label{sec:fftsec}
Figure~\ref{fig:strongscaling} shows the strong scaling performance of the FFT (right) and splits the runtime into computation and communication.
The library FFTW computes multiple FFTs and measures their execution times to find the optimal plan that achieves the best performance for each machine.
We use these optimal implementations.
Since the optimal plan selected by FFTW is based on a collection of \texttt{MPI\_Sendrecv}s, the results in Figure~\ref{fig:osuptp} are relevant to the following analysis.

FFT has a higher $Cm/Cp$ ratio than the N-Body simulation, and as expected, Figure~\ref{fig:strongscaling} shows that the communication overhead is much higher than in the previous application and can take more than $50\%$ of execution time.
There is a consistent spike in communication at $P{=}4$, which we suspect is due to implementation details of FFTW.
On all machines, the overall scaling of the FFT is sublinear, mainly due to communication overhead.
AWS R5 is the fastest platform, both in terms of total and communication time.
It is followed by Cori Haswell.
The computation times of Cori KNL and AWS C5 are comparable, but AWS C5 has a higher communication overhead, making it the slowest platform in this benchmark.

Despite comparable performance for point-to-point communication (Figure~\ref{fig:osuptp}), the cloud instances exhibit different performance for all processes on the node involved in the communication (Figure~\ref{fig:osu-alltoall}).
AWS C5 exhibits significantly worse performance for the \texttt{MPI\_Alltoall} benchmark, which explains the difference in communication performance between the two AWS instances for the FFT results (Figure~\ref{fig:strongscaling}).
AWS R5 is optimized for memory-intensive workloads, while AWS C5 is optimized for compute-intensive workloads.
Moreover, AWS C5 uses Amazon's in-house EFA interconnect, whose advertised bandwidth is $3\times$ lower than R5's.
Our hypothesis is that as the number of processes increases, the C5 interconnect is more subject to contention than R5's network.

AWS R5 is the best performing platform in this benchmark, as one would expect based on the results of our microbenchmarks and workload characterization.
The communication time on AWS R5 is comparable to or even lower than that on HPC systems.
Thus, it is not only the newer processor that contributes to the high performance for this application, but also the interconnect speed.
Previous literature has shown that FFTs for cloud instances have significantly lower performance than for HPC systems. 
The Magellan report~\cite{yelick2011magellan} describes the FFT as $4$ to $20\times$ slower than the HPC systems considered, running on 8 processes per node and $P{=}8$.
Our result is an important validation of the recent advances that cloud computing has made in networking technology to close the performance gap with HPC.

%% file: sec/conclusions.tex
\section{Conclusions and Future Work}\label{sec:conclusion}

Our work investigated the performance gap between current HPC and cloud computing systems to understand the nature of their differences and guide the design of future cloud systems.
In this work, we analyzed the cross-stack performance, from single core compute power, to memory subsystem, inter-node communication performance, and overall application performance.

In particular, we highlight that cloud computing can offer a greater variety of hardware configurations and newer technology due to continuous procurement cycles. If a study requires the latest technology or a particular memory size and processor type, these are more likely to be available in the cloud, while a given HPC system may offer only one or a small set of standardized resources suitable for typical scientific applications.
Our results contradict earlier findings on cloud interconnects, namely that networks for HPC instances within the cloud have improved to the point of providing competitive performance to that of HPC systems at modest scales.

On the other hand, cloud policies can limit what is available within an HPC cloud offering, e.g., one may need to make a request to the vendor to use more than a few instances, and the latest node architectures may not be available with the fast network.  
In contrast, in traditional HPC systems, the entire system typically has the same network, whose performance is mostly determined by the age of the system, as the procurement cycles are typically longer.

Our results showed that the compute and memory subsystem performance of cloud instances is competitive with HPC systems.
This is consistent with historical results demonstrating cloud competitiveness for compute-dominated workloads.

Cloud systems offered higher bandwidth and lower latency than HPC systems for point-to-point communication.
In the FFT benchmark, which is bisection-bandwidth limited, the performance of the compute-optimized cloud platform dropped, possibly due to network contention, while the platform optimized for memory-intensive applications significantly outperformed all other machines.
This represents a significant advance in cloud computing technology, as the performance of multinode FFT applications on HPC systems has historically been better than on cloud systems~\cite{yelick2011magellan}.
A larger scale performance study focusing on machine balance would be an interesting future work to analyze the gap on a larger scale.

Our work shows that today's cloud computing can provide competitive performance to HPC, not only for compute-intensive applications, but also on memory- and communication-intensive workloads.
The recent performance improvements of cloud instances may be due to the increasing demands of deep learning~\cite{jouppi2020domain, hazelwood2018applied}, potentially benefiting seemingly unrelated computational science as a byproduct. 
It is worth noting that our study focused on one cloud provider and it would be important to replicate the study on other providers to draw more generalized conclusions.
Given our results, an important future work would be a comparison focusing on elasticity and resource management, which together with our results would allow users to make informed decisions about which system is better suited for their applications.